\documentstyle[aas2pp4]{article}

\lefthead{Shimasaku}
\righthead{Clusters}

\begin{document}


\newcommand\vs{\vspace{10pt}}
\newcommand\hs{\hspace{5pt}}
\newcommand\ite{\par\hang\textindent}
\newcommand\iteite{\par\indent \hangindent2\parindent \textindent}
\newcommand\refe{\hangindent=50pt \hangafter=1 \noindent}
\def\la{\mathrel{\mathchoice {\vcenter{\offinterlineskip\halign{\hfil
$\displaystyle##$\hfil\cr<\cr\sim\cr}}}
{\vcenter{\offinterlineskip\halign{\hfil$\textstyle##$\hfil\cr
<\cr\sim\cr}}}
{\vcenter{\offinterlineskip\halign{\hfil$\scriptstyle##$\hfil\cr
<\cr\sim\cr}}}
{\vcenter{\offinterlineskip\halign{\hfil$\scriptscriptstyle##$\hfil\cr
<\cr\sim\cr}}}}}
\def\ga{\mathrel{\mathchoice {\vcenter{\offinterlineskip\halign{\hfil
$\displaystyle##$\hfil\cr>\cr\sim\cr}}}
{\vcenter{\offinterlineskip\halign{\hfil$\textstyle##$\hfil\cr
>\cr\sim\cr}}}
{\vcenter{\offinterlineskip\halign{\hfil$\scriptstyle##$\hfil\cr
>\cr\sim\cr}}}
{\vcenter{\offinterlineskip\halign{\hfil$\scriptscriptstyle##$\hfil\cr
>\cr\sim\cr}}}}}
\let\lsim=\la
\let\gsim=\ga

\newcommand\astroph{astro-ph}

\newcommand\kms{{\rm km} \ {\rm s}^{-1}}
\newcommand\Mpc{{\rm Mpc}}
\newcommand\oMpc{{\rm Mpc}^{-1}}
\newcommand\Lstar{L_\star}
\newcommand\Mgas{M_{\rm gas}}
\newcommand\Mstar{M_\star}
\newcommand\Msolar{M_\odot}
\newcommand\ergs{{\rm erg} \ {\rm s}^{-1}}
\newcommand\rhob{\rho^{\rm B}}
\newcommand\rhobcl{\rho^{\rm B}_{\rm cl}}
\newcommand\rhogcl{\rho^{\rm gas}_{\rm cl}}
\newcommand\Omegab{\Omega^{\rm B}}
\newcommand\Omegagcl{\Omega^{\rm gas}_{\rm cl}}
\newcommand\fb{f_{\rm B}}
\newcommand\deltac{\delta_{\rm c}}
\newcommand\rhodhcl{\rho^{\rm DH}_{\rm cl}}
\newcommand\rhocr{\rho^{\rm cr}}
\newcommand\fdh{f_{\rm M}}
\newcommand\Mb{M_{\rm B}}
\newcommand\Omeganu{\Omega_\nu}
\newcommand\Omegacdm{\Omega_{\rm CDM}}

\newcommand\cobe{{\it COBE}\hspace{5pt}}
\newcommand\rosat{{\it ROSAT}\hspace{5pt}}
\newcommand\rosate{{\it ROSAT}}
\newcommand\iras{{\it IRAS}\hspace{5pt}}

\newcommand\Qrms{Q_{{\rm rms}-PS}}


\title{Measuring the Density Fluctuation 
       From the Cluster Gas Mass Function}

\author{Kazuhiro Shimasaku}
\affil{Department of Astronomy, 
       School of Science, University of Tokyo,
       Bunkyo-ku, Tokyo 113, Japan}

\and

\affil{Research Center for the Early Universe, 
       School of Science, University of Tokyo,
       Bunkyo-ku, Tokyo 113, Japan}


\begin{abstract}
We investigate the gas mass function of clusters of galaxies  
to measure the density fluctuation spectrum on cluster scales. 
The baryon abundance confined in rich clusters 
is computed from the gas mass function and
compared with the mean baryon density in the universe which is 
predicted by the Big Bang Nucleosynthesis.
This baryon fraction and the slope of 
the gas mass function put constraints on $\sigma_8$, 
the rms linear fluctuation on scales of $8h^{-1}\Mpc$, and 
the slope of the fluctuation spectrum, 
where $h$ is the Hubble constant in units of 100 $\kms \oMpc$.
We find $\sigma_8 = 0.80 \pm 0.15$ 
and $n \sim -1.5$ for $0.5 \le h \le 0.8$,
where we assume that the density spectrum is approximated by
a power law on cluster scales: 
$\sigma(r) \propto r^{-{3+n\over{2}}}$.
Our value of $\sigma_8$ is independent of 
the density parameter, $\Omega_0$, and
thus we can estimate $\Omega_0$ by combining 
$\sigma_8$ obtained in this study with those from 
$\Omega_0$-dependent analyses to date.
We find that $\sigma_8(\Omega_0)$ derived from the cluster 
abundance such as the temperature function gives $\Omega_0 \sim 0.5$ 
while $\sigma_8(\Omega_0)$ measured from the peculiar velocity 
field of galaxies gives $\Omega_0 \sim 0.2-1$, depending on the 
technique used to analyze peculiar velocity data.
Constraints are also derived for 
open, spatially flat, and tilted Cold Dark Matter models 
and for Cold $+$ Hot Dark Matter models.
\end{abstract}



\section{Introduction}

To measure the spectrum of density fluctuations is 
one of the key issues in discussing the structure formation in the 
universe.
Clusters of galaxies are suitable objects to measure the 
spectrum on scales of $\sim 10 h^{-1}$ Mpc, where $h$ is 
the Hubble constant in units of 100 $\kms \oMpc$.
This is because the abundance of clusters is sensitive to the nature  
of the spectrum, in particular the amplitude, 
and because the fluctuations on cluster scales can be 
reliably discussed by linear theory.

Henry \& Arnaud (1991) derived $\sigma_8 = 0.59 \pm 0.02$ 
and $n= -1.7^{-0.65}_{+0.35}$ from the X-ray temperature 
function of clusters for $\Omega_0=1$ universes,
where $\sigma_8$ is the rms linear fluctuation 
on scales of $8h^{-1}\Mpc$, 
$\Omega_0$ is the cosmological density parameter, 
and they assumed that the density spectrum is approximated by
a power law in wavenumber as $P(k) \propto k^n$ 
on cluster scales.
White, Efstathiou, \& Frenk (1993a) obtained
$\sigma_8 \simeq (0.57 \pm 0.05) \Omega_0^{-0.56}$ 
using the spatial number density of rich clusters.
Similar results for $\sigma_8$ were also obtained by other authors 
for open and spatially flat Cold Dark Matter (CDM) models
using Henry \& Arnaud's (1991) temperature function data;
Eke, Cole, \& Frenk (1996) found 
$\sigma_8 = (0.50 \pm 0.04) \Omega_0^{-\alpha}$ 
and Viana \& Liddle (1996) found $\sigma_8 = 0.6 \Omega_0^{-\alpha}$;
in both estimates $\alpha$ varies from $\sim 0.4$ to $\sim 0.6$  
depending on the dark matter model assumed.
There is, however, a problem with those measurements of $\sigma_8$ 
that one cannot know $\sigma_8$ unless $\Omega_0$ is given;
in general, the measured quantity is not $\sigma_8$ but 
a combination of $\sigma_8$ and $\Omega_0$  
like $\sigma_8 \Omega_0^{0.6}$.  
This problem is common to almost all methods for measuring 
$\sigma_8$, such as the one using
the peculiar velocity field of galaxies.

In this paper, we measure observationally 
the fraction of baryons confined in 
clusters of galaxies to the total baryons in the universe  
using the cluster gas mass function, and 
give $\Omega_0$-independent measurements of $\sigma_8$ 
by comparing the observed baryon fraction 
with the theoretical prediction.
As will be seen in \S$\ $3, 
the theoretical derivation of the baryon fraction does not 
depend on $\Omega_0$,
since the baryon fraction measures essentially 
the fraction of density fluctuations with an overdensity 
larger than the critical value. 
We find $\sigma_8 = 0.80 \pm 0.15$, with the quoted errors 
including the uncertainties in $h$ ($0.5 \le h \le 0.8$).
The power-law index of the spectrum, $n$, is also measured.

The plan of this paper is as follows.
In \S$\ $2 we derive the cluster gas mass function
from the observed sample 
and compute the fraction of baryons confined in rich clusters.
In \S$\ $3, we describe an analytic model of gravitational 
halo formation to predict the fraction of mass 
confined in dark haloes with a given mass range.
In \S$\ $4, we estimate $\sigma_8$ and $n$. 
We also derive constraints on three sets of CDM models 
and a set of Cold $+$ Hot Dark Matter (CHDM) models in \S$\ $4.
Finally, \S$\ $5 summarizes our conclusions.


\section{Data}

In this section, we first derive the gas mass function
of clusters from X-ray data, and then estimate 
the fraction of baryons confined in 
clusters with a given gas mass range.
Ebeling et al. (1996a) constructed an X-ray luminosity function 
(XLF) of clusters of galaxies from the \rosat Brightest Cluster 
Sample, which is an X-ray selected, flux limited sample of 172 
clusters compiled from \rosat All-Sky Survey data.
They found that the XLF is well fitted by a Schechter function:
\begin{equation}
\phi(L) dL = A L^{-\alpha} \exp \left( -{L \over{\Lstar}} \right) dL 
\end{equation}
\noindent
with $A= (4.58 \pm 0.76) \times 10^{-6}\ h^3\ {\rm Mpc}^{-3}
\times (0.25 \times 10^{44}\ h^{-2}\ \ergs)^{\alpha-1}$,
$\Lstar = (2.23 \pm 0.42) \times 10^{44}\ h^{-2}\ \ergs$, 
and $\alpha = 1.78 \pm 0.09$.
Here, $L$ is the X-ray luminosity in the $0.1 - 2.4$ keV band.

Ebeling et al. (1996b) presented an X-ray sample of 242 Abell 
clusters from \rosat All-Sky Survey data in which X-ray 
luminosity in the $0.1 - 2.4$ keV band is tabulated.
Among those, the gas mass 
within the Abell (1958) radius ($\equiv 1.5 h^{-1}$ Mpc) 
is given for 40 clusters 
in Jones \& Forman (1984) and Arnaud et al. (1992).
The Abell radius is close to the typical 
virialization radius of rich clusters, and thus 
it is reasonable to suppose that most of the cluster gas 
is within the Abell radius. 
From the luminosity and the gas mass 
data of the forty clusters, we find 
$$
\Mgas = (0.34 \pm 0.04) 
\hspace{200pt}
$$
\begin{equation}
\times \left({L\over{10^{44}}}\right)^{0.65 \pm 0.07} 
(h^{-2.5} 10^{14} \Msolar),
\end{equation}
\noindent
where $L$ is in units of $h^{-2}\ \ergs$ and 
$\Mgas$ is the gas mass within the Abell radius.
When estimating the errors in the coefficients of 
the $L$-$\Mgas$ relation (eq.[2]),
we assume that the observed $\Mgas$ for each cluster 
has an error of $\sigma_{\log \Mgas} = 0.25$, 
which is the typical scatter of the data points 
around equation (2).
Figure 1 shows the distribution of the forty clusters 
on the $L$ - $\Mgas$ plane.

\begin{figure}
\plotone{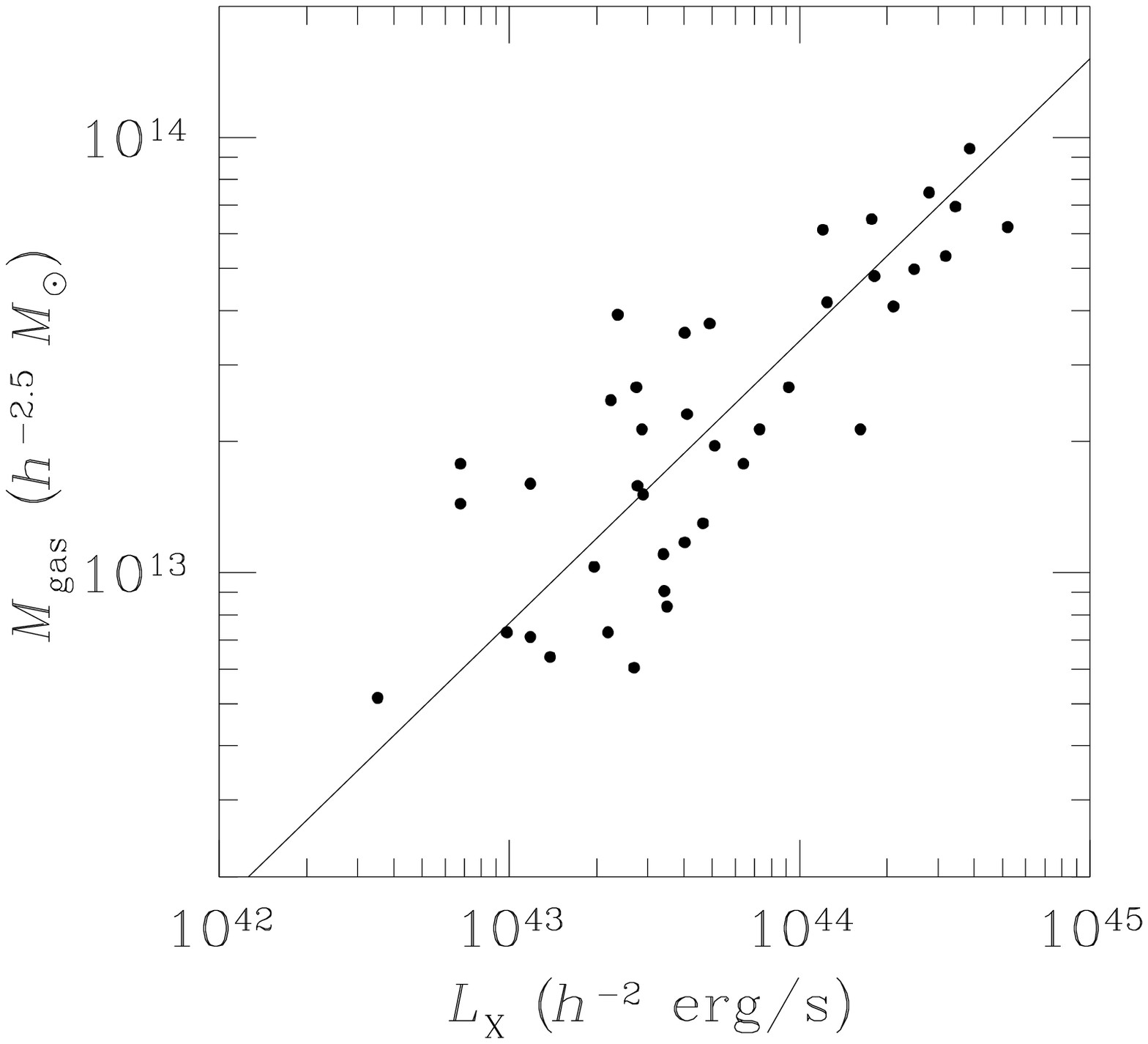}
\caption
{$L$ - $\Mgas$ relation of forty clusters.
Equation (2) is shown as the solid line. 
\label{fig1}}
\end{figure}
 
From the XLF and the $L$ - $\Mgas$ relation, we obtain 
the gas mass function:
$$
\psi(\Mgas) d \Mgas
\hspace{200pt} {\ } 
$$
\vspace{-10pt}
$$
= (1.28 \pm 0.46) \times 10^{-6}
\times \left( {\Mgas \over{\Mstar}} \right)^{-2.20 \pm 0.15}
\hspace{160pt} {\ } 
$$
\vspace{-5pt}
\begin{equation}
\times \exp 
  \left[ -\left({\Mgas \over{\Mstar}}\right)^{1.54 \pm 0.02} \right] 
d \left( {\Mgas \over{\Mstar}} \right)
(h^3\ {\rm Mpc}^{-3}),
\end{equation}
\noindent
where $\Mstar = 0.57 \times 10^{14} h^{-2.5} \Msolar$.
This gas mass function is consistent with 
the (cumulative) gas mass function derived by 
Burns et al. (1996) using optically selected poor and rich clusters.

We now compute the fraction of baryons confined in rich clusters.
This calculation consists of three steps:
(1) derive the gas density in the universe 
contributed from clusters, $\rhogcl$, from equation (3);
(2) add to $\rhogcl$ the contribution from the stellar mass 
of cluster galaxies to obtain the baryon density 
contributed from clusters, $\rhobcl$;
(3) compute the baryon fraction 
$\fb \equiv \rhobcl/\rhob_0$, where $\rhob_0$ is 
the mean baryon density in the universe  
predicted by the Big Bang Nucleosynthesis.  

First, we compute $\rhogcl$ by 
\begin{equation}
\rhogcl = \int_{M_1}^{M_2} M \psi (M) dM,
\end{equation}
\noindent
where $M_1$ and $M_2$ are the lower and upper limits 
of cluster gas mass.
Since the XLF and $L$ - $\Mgas$ relation are available 
over $10^{42} \lsim L \lsim 10^{45}$ and 
$10^{43} \lsim L \lsim 10^{45}\ h^{-2}\ \ergs$, respectively,
we set 
$(M_1, M_2) = (7.6 \times 10^{12}, 1.5 \times 10^{14})$
($h^{-2.5} \Msolar$),
which corresponds to
$(L_1, L_2) = (10^{43}, 10^{45})$ ($h^{-2}\ \ergs$).
Performing the integration in equation (4)
\footnote{Actually, we calculate 
$\int_{L_1}^{L_2} M(L)\phi(L)dL$ in order to 
evaluate the errors in $\rhogcl$ more easily.}, 
we obtain
\begin{equation} 
\rhogcl = (1.51 \pm 0.32) \times 10^8\ h^{0.5}\ \Msolar \ \Mpc^{-3},
\end{equation} 
\noindent
which is translated into the density parameter 
$\Omegagcl = (5.45 \pm 1.17) \times 10^{-4}\ h^{-1.5}$.
This value of $\Omegagcl$ agrees roughly with
what Persic \& Salucci (1992) obtained 
for a somewhat narrower mass range of clusters,
$\Omegagcl \simeq 3.5 \times 10^{-4}\ h^{-1.3}$, from 
an analysis using another XLF derived before \rosate.

We fit the observed gas mass function by a power law 
over the range of $M_1$ and $M_2$ to find 
\begin{equation}
\psi(M) \propto M^{-3.3 \pm 0.2}.
\end{equation}
\noindent
This power-law index will also be used to put constraints 
on the density fluctuation spectrum, in particular its slope.

Second, we add the contribution from the stellar mass of 
galaxies in clusters to $\rhogcl$.
According to White et al. (1993b), the mass ratio of the stellar 
component to the hot gas is $0.2 h^{1.5}$ 
for the Coma Cluster.
Since Coma is a typical rich cluster, we assume that other 
clusters have the same ratio.
Under this assumption, the density of baryons in clusters, 
$\rhobcl$, is calculated by 
\begin{equation} 
\rhobcl = \rhogcl (1 + 0.2 h^{1.5}).
\end{equation} 

Finally, let us derive $\fb$.
The density parameter of baryons in the universe, $\Omegab_0$, 
is estimated using the Big Bang Nucleosynthesis.
We take Walker et al.'s (1991) estimate:
$\Omegab_0 h^2 = 0.0125 \pm 0.0025$.
We do not adopt recent estimates based on 
the deuterium abundance in QSO absorption systems, 
because they are still quite controversial
({\it e.g.}, Carswell et al. 1996; Rugers \& Hogan 1996; 
Tytler, Fan, \& Burles 1996; Wampler 1996).
Since $\rhob_0 = \rhocr_0 \Omegab_0$ 
($\rhocr_0 \equiv 3 H_0^2/8 \pi G$ 
is the critical matter density and $H_0$ is the Hubble constant), 
we obtain 
\begin{equation} 
\fb = (0.044 \pm 0.013) (1 + 0.2 h^{1.5}) h^{0.5}.
\end{equation} 
\noindent
The errors are at $1\sigma$ levels and 
include both the errors in equation (5) and in $\Omegab_0$.


\section{Model}

In this section we derive analytically 
the fraction of matter confined in haloes with a given 
mass range, based on the gravitational halo formation model.
The Press-Schechter (1974) theory predicts the number 
density of haloes with a comoving radius $r$ as
$$
n(r) dr 
= - {3\over{4\pi}} \left( {2\over{\pi}}\right)^{1\over{2}}
   \deltac 
\hspace{200pt}
$$
\vspace{-10pt}
\begin{equation}
\times {1\over{r^3}} {1\over{\sigma^2(r)}}
   {d\sigma(r) \over{dr}}
   \exp \left[ - {\deltac^2\over{2 \sigma^2(r)}}\right]
   dr,
\end{equation}
\noindent
where $\deltac$ is the critical linear overdensity and 
$\sigma(r)$ is the rms of the linear density fluctuations 
in top-hat windows of radius $r$; 
$\sigma_8$ is defined as $\sigma(8h^{-1}\Mpc)$.
The fraction of matter in the universe 
which is confined in haloes with $r_1 \le r \le r_2$ is expressed as 
\begin{equation}
\fdh = { \int_{r_1}^{r_2} M(r) n(r) dr 
     \over{ \rho_0} }.
\end{equation}
\noindent
Here, $M(r)$ is the total mass of the halo with a comoving radius $r$ 
and $\rho_0 \equiv \rhocr_0 \Omega_0$. 
We assume that the total mass of haloes is within the Abell radius.
Since 
\begin{equation}
M(r) = {4\pi \over{3}} \rho_0 r^3,
\end{equation}
equation (10) is rewritten as
\begin{equation}
\fdh = \int_{r_1}^{r_2} {4\pi\over{3}} r^3 n(r) dr.
\end{equation}

We then assume that the baryon-to-total mass ratio 
of haloes is equal to $\rhob_0/\rho_0$. 
This assumption is reasonable 
because $N$-body simulations to date have suggested that 
hydrodynamical effects do not significantly influence
the baryon-to-total mass ratio within the Abell radius 
({\it e.g.}, White et al. 1993b;
Lubin et al. 1996 and references therein). 
Thus, a cluster with a baryon mass $\Mb$ have 
a total mass $M_{\rm tot} = \Mb \rho_0/\rhob_0$, and 
thus have a comoving radius: 
\begin{equation}
r = \left( {3\Mb \over{4\pi \rhob_0}} \right)^{1\over{3}}.
\end{equation}
\noindent
Equations (9), (12), and (13) show that 
$\fdh$, which is equal to $\fb$ by the assumption above, 
is computed if $\sigma(r)$, $\rhob_0$, and 
the lower and upper limits of $\Mb$ are given.
We take $\deltac=1.69$ throughout this paper for simplicity,
because $\deltac$ depends on 
$\Omega_0$ and the cosmological constant $\lambda_0$ 
only weakly ({\it e.g.}, Lilje 1992; Lacey \& Cole 1993).


\section{Results and Discussion}


\subsection{Estimates of the Amplitude and Shape of the 
Fluctuation}

In this subsection, 
we assume that the density fluctuation spectrum is a power law 
on cluster scales: 
\begin{equation}
\sigma(r) 
 = \sigma_8 \left( {r \over{8 h^{-1}\Mpc}} \right)^{-{3+n\over{2}}},
\end{equation}
\noindent
and estimate $\sigma_8$ and $n$ from $\fb$ and the slope of 
the gas mass function.

\begin{figure}
\plotone{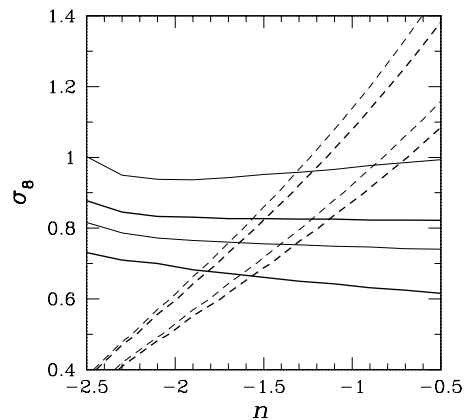}
\caption
{Constraints on $\sigma_8$ and $n$. 
The region between the thick (thin) solid lines 
indicates the range allowed by the baryon fraction $\fb$ 
for $h=0.5$ (0.8). 
The region between the thick (thin) dashed lines is for
the constraint for $h=0.5$ (0.8) derived 
from the slope of the gas mass function. 
\label{fig2}}
\end{figure}

Figure 2 shows the constraints on $\sigma_8$ and $n$.
Results are shown for two extreme values of $h$, 
since the observed and predicted baryon fractions 
depend on $h$ (see eqs.[8] and [13]).
The two values of $h$ taken here, $0.5$ and $0.8$, 
roughly correspond to the lower and upper limits of 
recent measurements, respectively (see, {\it e.g.,} Freedman 1996). 
The region between the two thick (thin) solid lines indicates
the range allowed by the baryon fraction, $\fb$, 
for $h=0.5$ (0.8); for each $h$, the region corresponds 
to the $1 \sigma$ range of $\fb$ (eq.[8]).
The region between the two thick (thin) dashed lines is for 
the constraint from the slope of the gas mass function 
for $h=0.5$ (0.8); 
for each $h$, the region corresponds to the $1 \sigma$ 
range of the slope (eq.[6]).
In deriving the constraint from the slope of the gas mass function, 
we fit the theoretical mass function 
with a power law, $n(M) \propto M^a$, 
over the range of $M$ corresponding to the gas mass range 
($M_1$ and $M_2$ in \S$\ $3), and compare $a$ with 
the power-law index of equation (6).
We see in Figure 2 two features common to the $h=0.5$ and 0.8 cases.
First, $\fb$ places almost the unique constraint on $\sigma_8$ 
irrespective of $n$.
This is because $8\ h^{-1}\Mpc$ is the typical comoving radius 
of rich clusters.
Second, a higher $\sigma_8$ is required for a higher $n$ 
to reproduce the observed slope of the gas mass function. 

From Figure 2 
we find the allowed values of $\sigma_8$ and $n$ to be  
$(\sigma_8,n) = (0.75 \pm 0.1, -1.5 \pm 0.4)$ for $h=0.5$, 
and $(0.85 \pm 0.1, -1.3 \pm 0.4)$ for $h=0.8$;
a lower $h$ gives a lower $\sigma_8$.
Adopting $0.5 \le h \le 0.8$ as the observational uncertainty in $h$,
we can express the range of $\sigma_8$ by
\begin{equation}
\sigma_8 = 0.80 \pm 0.15.
\end{equation}

In the rest of this subsection, 
we compare $\sigma_8$ obtained above with those 
from $\Omega_0$-dependent analyses to date,
and then estimate $\Omega_0$.
Henry \& Arnaud (1991) derived $\sigma_8 = 0.59 \pm 0.02$ 
and $n = -1.7^{-0.65}_{+0.35}$ 
from the cluster temperature function assuming $\Omega_0=1$. 
The value of $n$ they obtained 
agrees very well with what we obtain above.
White et al. (1993a) derived
$\sigma_8 \simeq (0.57 \pm 0.05) \Omega_0^{-0.56}$ 
using the cumulative abundance of rich clusters.
The basic procedures by Henry \& Arnaud (1991) 
and White et al. (1993a) to measure $\sigma_8$ 
are the same in the sense that both of them 
used the cluster abundance and applied 
the Press-Schechter (1974) theory.
Thus, White et al.'s (1993a) estimate of $\sigma_8$ 
for arbitrary $\Omega_0$ may be regarded as 
a generalization of Henry \& Arnaud's (1991).
Another method for measuring $\sigma_8$ is 
to use the peculiar velocity field of galaxies,
though a large inconsistency is seen among recent measurements.
Two representative measurements are those given in 
Kolatt \& Dekel (1997) and Willick et al. (1996).
Kolatt \& Dekel (1997) obtained 
$\sigma_8 = (0.71-0.77 [\pm 0.12]) \Omega_0^{-0.6}$ 
from the POTENT analysis 
(Dekel, Bertschinger, \& Faber 1990; Dekel 1994)
of the Mark III peculiar velocity catalog (Willick et al. 1997).
Willick et al. (1996) found 
$\sigma_8 = (0.34 \pm 0.05) \Omega_0^{-0.6}$,
applying another technique to the same catalog.
One can also measure $\sigma_8$ (as a function of $\Omega_0$)
using the redshift-space anisotropy of galaxy clustering;
examples are Fisher et al. (1994) and 
Cole, Fisher, \& Weinberg (1995), both of which derived 
similar values to what Willick et al. (1996) obtained
\footnote{The quantity Fisher et al. (1994) and Cole et al. (1995) 
measured is not $\sigma_8$ but 
$\beta_I \equiv \Omega_0^{0.6}/b_I$, 
where $b_I$ is the bias parameter of \iras galaxies. 
However, as is shown in Willick et al. (1996), 
one can derive $\sigma_8(\Omega_0)$ from $\beta_I$ 
and the observed rms fluctuations of \iras galaxies on 
an $8h^{-1}$ Mpc scale,
assuming that biasing is independent of scale.}.

Since our estimate of $\sigma_8$ is independent of $\Omega_0$ 
while those given in the previous papers listed above 
depend on $\Omega_0$ in a way like 
$\sigma_8(\Omega_0) = {\rm const.} \times \Omega_0^{-0.6}$,
we can compute $\Omega_0$ by comparing them. 
We take three estimates of $\sigma_8(\Omega_0)$ 
as representatives of the previous 
ones: those given in White et al. (1993a), Kolatt \& Dekel (1997), 
and Willick et al. (1996).
A comparison between our $\sigma_8$ and White et al.'s (1993a) 
$\sigma_8(\Omega_0)$ gives  
$\Omega_0 = 0.6^{+0.3}_{-0.2}$ (for $h=0.5$) and  
$\Omega_0 = 0.5^{+0.2}_{-0.2}$ ($h=0.8$).
From Kolatt \& Dekel's (1997) measurement we obtain 
$\Omega_0 = 1.0^{+0.7}_{-0.5}$ ($h=0.5$) and  
$\Omega_0 = 0.8^{+0.5}_{-0.3}$ ($h=0.8$).  
Finally, Willick et al.'s (1996) $\sigma_8(\Omega_0)$
gives $\Omega_0 = 0.3^{+0.2}_{-0.1}$ ($h=0.5$) and 
$\Omega_0 = 0.2^{+0.1}_{-0.1}$ ($h=0.8$).
We then compare these values with that computed from
the cluster baryon-to-total mass ratio and $\Omegab_0$. 
White et al. (1993b) found that the baryon-to-total mass ratio 
for the Coma Cluster is $0.009 + 0.05 h^{-1.5}$ 
(the $1 \sigma$ error at a given value of $h$ is $\sim 30\%$). 
Adopting this ratio as the global value in the universe 
and using $\Omegab_0=(0.0125 \pm 0.0025) h^{-2}$, 
we obtain $\Omega_0 = 0.33 \pm 0.1$ for $h=0.5$
and $\Omega_0 = 0.25 \pm 0.1$ for $h=0.8$.
These values of $\Omega_0$ are in very good agreement 
with those derived 
from Willick et al.'s (1996) measurement of $\sigma_8 (\Omega_0)$, 
but are smaller than those derived from 
White et al.'s (1993a) and Kolatt \& Dekel's (1997),
though the disagreements are less than $2\sigma$ levels.
To summarize, we find that there exists a region on 
the $\Omega_0$ - $\sigma_8$ plane 
where the three constraints overlap with each other 
within the observational uncertainties,
{\it i.e.}, the constraints on $\sigma_8$ obtained from 
the gas mass function in this paper, 
on $\sigma_8 (\Omega_0)$ derived from the 
$\Omega_0$-dependent analyses discussed above, 
and on $\Omega_0$ derived from the cluster baryon-to-total 
mass ratio and $\Omegab_0$.  


\subsection{Constraints on Cosmological Models}

In this subsection, we examine four sets of 
cosmic structure formation models including 
(1) open ($\lambda_0=0$) CDM models with $\Omega_0$ the free 
parameter,
(2) spatially flat CDM models with $\Omega_0$ the free parameter,
(3) tilted CDM models with $n$ the free parameter, and
(4) CHDM models with the density parameter of the hot dark 
matter, $\Omeganu$, the free parameter. 
In what follows, $n$ denotes the power-law index of the primordial 
({\it i.e.}, before filtered by the transfer function) density 
fluctuation spectrum, not the index of the current 
spectrum on cluster scales.
Table 1 summarizes the model parameters.
For each set, we constrain the above free parameter by requiring 
that it matches simultaneously the observed baryon fraction 
and the 4-year \cobe DMR data (Bennett et al. 1996).
We take the functional forms of the CDM and CHDM spectra
from those given in Kitayama \& Suto (1996) 
and Ma (1996), respectively 
(these authors took the form of the CDM spectrum 
from Bardeen et al. 1986 and corrected it for the baryon component). 
We do not use the slope of the gas mass function 
to constrain the models
because the CDM and CHDM spectra on cluster scales 
give power-law indices which are close to those derived 
in \S$\ $4.1.
We assume $\Omega_0=1$ for the tilted CDM models and the CHDM models 
for definiteness, although 
this assumption is apparently inconsistent with 
the values of $\Omega_0$ derived from $\Omegab_0$ 
and the baryon-to-total mass ratio for Coma (see \S$\ $4.1).

\vspace{-5pt}
\begin{table}
\tabcolsep 1.5pt
\caption{Model parameters.}
\begin{tabular}{cccccc}
\hline
\hline
Model & $h$ & $n$ & $\Omega_0$ & $\lambda_0$ & $\Omeganu$ \\
\hline
open CDM           & $0.5,0.8$ & $1$ & varying & $0$          & $0$ \\
spatially flat CDM & $0.5,0.8$ & $1$ & varying & $1-\Omega_0$ & $0$ \\
tilted CDM         & $0.5$     & varying & $1$ & $0$          & $0$ \\
CHDM               & $0.5$     & $1$ & $1$     & $0$      & varying \\
\hline
\end{tabular}
\label{tab1}
\end{table}

\begin{figure}
\plotone{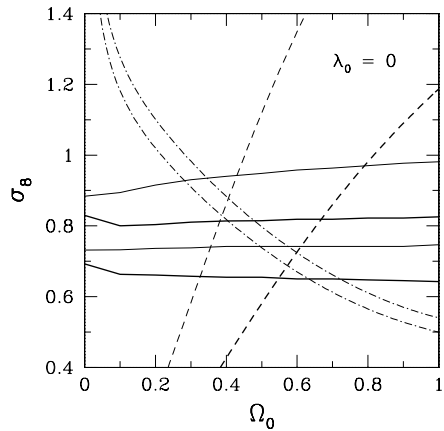}
\caption
{Constraints on $\sigma_8$ and $\Omega_0$ for open CDM models. 
The region between the thick (thin) solid lines 
indicates the range allowed by the baryon fraction $\fb$ 
for $h=0.5$ (0.8). 
The value of $\sigma_8$ given by the 4-year \cobe data 
for $h=0.5$ (0.8) is shown by the thick (thin) dashed line.
The region between the dash-dotted lines 
indicates the constraint derived by Kitayama \& Suto (1997).
\label{fig3}}
\end{figure}

\begin{figure}
\plotone{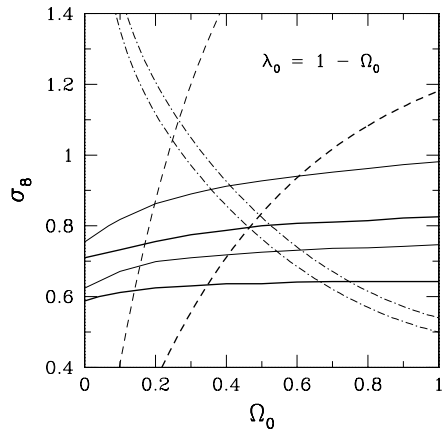}
\caption
{Same as Figure 4, but for spatially flat CDM models.
\label{fig4}}
\end{figure}

\begin{figure}
\plotone{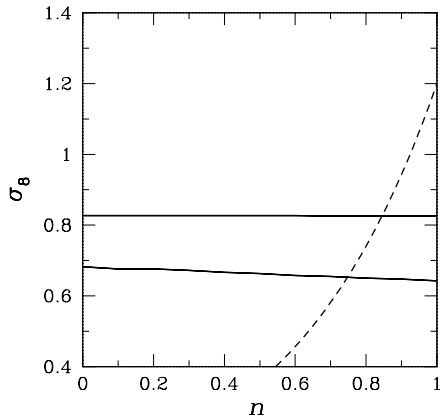}
\caption
{Constraints on $\sigma_8$ and $n$ for tilted CDM models. 
The region between the thick solid lines 
indicates the range allowed by the baryon fraction $\fb$.
The value of $\sigma_8$ given by the 4-year \cobe data 
is shown by the thick dashed line.
\label{fig5}}
\end{figure}

\begin{figure}
\plotone{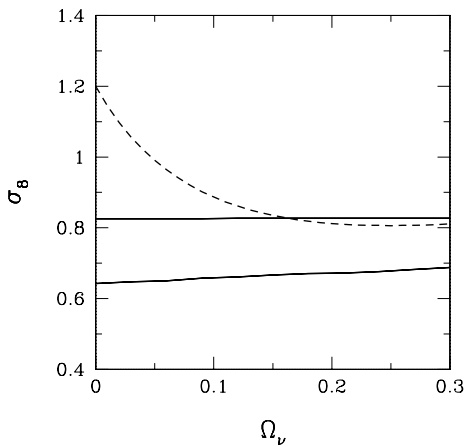}
\caption
{Constraints on $\sigma_8$ and $\Omeganu$ for CHDM models. 
The meaning of the curves is the same as in Figure 5.
\label{fig6}}
\end{figure}

\vs
\noindent
{\it Open CDM models}

Figure 3 plots the constraints on $\sigma_8$ and $\Omega_0$ 
for open CDM models.
Results are shown for two extreme values of $h$. 
The region between the two thick (thin) solid lines is  
allowed by the baryon fraction, $\fb$, for $h=0.5$ (0.8).
The thick (thin) dashed line indicates $\sigma_8$ for $h=0.5$ (0.8) 
which is estimated from the 4-year \cobe data (Bennett et al. 1996);
we use the formulae given in Bunn \& White (1996) 
to derive $\sigma_8$ for open and spatially flat CDM models.
For $h=0.5$, $\Omega_0 \simeq 0.55 - 0.65$ is required to reproduce 
simultaneously the baryon fraction and the \cobe data.
For $h=0.8$, the allowed range is $\Omega_0 \simeq 0.35-0.45$.
In any case, $\Omega_0=1$ models are disfavored irrespective of $h$.
The allowed values of $\Omega_0$ obtained here agree with 
those recently obtained for scale-invariant 
CDM models using the cluster abundance;
Bahcall \& Cen (1993) found that low-density 
($\Omega_0 \sim 0.25-0.35$), low-bias ($\sigma_8 \sim 1 - 0.8$) 
CDM models, with or without $\lambda_0$, 
are consistent with the observed cluster mass function;
Kitayama \& Suto (1996) found that if $\Omega_0 \sim 0.2-0.5$ 
and $h=0.7$ are taken, CDM models with the fluctuation spectrum 
normalized by the \cobe data reproduce the observed cluster 
temperature function irrespective of $\lambda_0$; 
Liddle et al.(1996) concluded that for $h>0.6$, 
open CDM models for which the fluctuation spectrum is normalized 
by the \cobe data match the observations of galaxies, 
clusters, and damped Ly$\alpha$ systems considered in 
their paper only for the range $0.35 < \Omega_0 < 0.55$;
Kitayama \& Suto (1997) derived constraints on $\sigma_8$ 
and $\Omega_0$ for CDM models from the observed 
$\log N$ - $\log S$ relation of X-ray clusters.
As an example we plot the result obtained by 
Kitayama \& Suto (1997) in Figure 3; 
the region between the dash-dotted lines 
indicates the allowed range obtained by Kitayama \& Suto (1997) 
\footnote{Kitayama \& Suto's (1997) result plotted here 
is for $h=0.7$, since they gave an analytic formula for the 
constraint on $\sigma_8$ and $\Omega_0$ only for $h=0.7$ models.
However, figure 3 in their paper shows that 
the result is robust against varying $h$ from 0.5 to 0.8.}.
We find that for each value of $h$, there is an area 
on the $\Omega_0$ - $\sigma_8$ plane where the model reproduces 
both the baryon fraction, the $\log N$ - $\log S$ relation, 
and the \cobe data.
The constraints on $\Omega_0$ obtained here are also roughly
consistent with those derived from the cluster baryon-to-total 
mass ratio and $\Omegab_0$ in \S$\ $4.1. 

\vs
\noindent
{\it Spatially flat CDM models}

Figure 4 shows the constraints on $\sigma_8$ and $\Omega_0$ 
for spatially flat CDM models.
The meaning of the lines is the same as in Figure 3.
For $h=0.5$, $\Omega_0 \simeq 0.35-0.45$ is required, while
as low as $\Omega_0 \simeq 0.15 - 0.2$ is required for $h=0.8$. 
These values of $\Omega_0$
are also consistent with the previous estimates mentioned above.
As an example we compare our results with that
derived by Kitayama \& Suto (1997) in Figure 4.
For $h=0.5$, models with $(\Omega_0, \sigma_8) \simeq (0.45, 0.8)$
match both the baryon fraction, 
the $\log N$ - $\log S$ relation, and the \cobe data, 
though no such set of $(\Omega_0, \sigma_8)$ is found for $h=0.8$.

\vs
\noindent
{\it Tilted CDM models}

Figure 5 shows the results for tilted CDM models.
We fix $h = 0.5$ so that the age of the universe is 13 Gyr 
and is consistent with that predicted from globular clusters 
with $(1-2)\sigma$ levels.
The region between the two solid lines indicates the constraint 
from the baryon fraction.
The dashed line shows $\sigma_8$ which is
estimated from the 4-year \cobe data (Ma 1996).
We find that $n \simeq 0.75-0.85$ is required 
to match both the baryon fraction and the \cobe data.
Liddle et al. (1995) found that $h=0.5$ tilted models 
for which the spectrum is normalized by the \cobe data 
seem to be marginally consistent with the observations 
considered in their paper if $n \sim 0.6-0.7$ is taken, 
which is slightly smaller than the value we obtain here.

\vs
\noindent
{\it CHDM models}

Finally, CHDM models are examined in Figure 6.
The meaning of the lines is the same as in Figure 5.
We find that $\Omeganu \gsim 0.15$ is required 
to reproduce the baryon fraction and the \cobe data.
This result is consistent with what Liddle et al. (1995) obtained;
they found that when $h=0.5$ and $n=1$ are taken,
$\Omeganu \sim 0.2-0.3$ models 
with the fluctuation spectrum normalized by the \cobe data  
match the observation. 


\section{Conclusions}

We have constructed observationally the gas mass function of 
clusters of galaxies to measure the fraction of baryons 
confined in clusters to the total baryons in the universe.
Comparing this baryon fraction and the slope of the gas mass 
function with the prediction by the gravitational halo formation 
model, we have found $\sigma_8 = 0.80 \pm 0.15$ 
and $n \sim -1.5$ for $0.5 \le h \le 0.8$.
Our value of $\sigma_8$ is independent of $\Omega_0$, 
and thus we can estimate $\Omega_0$ from
the present result and previous ones in which  
$\sigma_8$ was obtained as a function of $\Omega_0$.
We have found that $\sigma_8(\Omega_0)$ derived from 
the cluster abundance gives $\Omega_0 \sim 0.5$ 
while $\sigma_8(\Omega_0)$ measured from 
the peculiar velocity field of galaxies gives 
$\Omega_0 \sim 0.2-1$, depending on the technique 
used to analyze the peculiar velocity data.
We have also examined four sets of cosmic structure formation 
models and have found that the following models match
both the observed baryon fraction and the \cobe data;
open CDM models with $\Omega_0 \simeq 0.55-0.65$ (for $h=0.5$) 
and $\Omega_0 \simeq 0.35-0.45$ ($h=0.8$); 
spatially flat CDM models with $\Omega_0 \simeq 0.35 - 0.45$ ($h=0.5$) 
and $\Omega_0 \simeq 0.15 - 0.2$ ($h=0.8$);
tilted CDM models with $n \simeq 0.75-0.85$;
CHDM models with $\Omeganu \gsim 0.15$.


\acknowledgments

We thank Yasushi Suto and Tetsu Kitayama for a useful 
discussion on the Press-Schechter theory and 
for valuable suggestions.
This research was supported in part by 
the Grants-in-Aid by the Ministry of Education, 
Science, Sports and Culture of Japan (07CE2002) 
to RESCEU (Research Center for the Early Universe).



\clearpage

\begin{thebibliography}{}

\bibitem[Abell 1958]{}
Abell, G. O. 1958, \apjs, 3, 211

\bibitem[Aunaud et al. 1992]{}
Arnaud, M., Rothenflug, R., Boulade, O., Vigroux, L., 
\& Vangioni-Flam, E. 1992, \aap, 254, 49

\bibitem[Bahcall \& Cen 1993]{}
Bahcall, N. A. \& Cen, R. 1993, \apj, 407, L49

\bibitem[Bardeen et al. 1986]{}
Bardeen, J. M., Bond, J. R., Kaiser, N., \& Szalay, A. S. 
1986, \apj, 304, 15

\bibitem[Bennett et al. 1996]{}
Bennett, C. L., et al. 1996, \apj, 464, L1

\bibitem[Bunn \& White 1996]{}
Bunn, E. F. \& White M. 1996, \astroph/9607060

\bibitem[Burns et al. 1996]{}
Burns, J. O. et al. 1996, \apj, 467, L49

\bibitem[Carswell et al 1996]{}
Carswell, R. F. et al. 1996, \mnras, 278, 506

\bibitem[Cole, Fisher, \& Weinberg 1995]{}
Cole, S., Fisher, K. B., \& Weinberg, D. H. 
1995, \mnras, 275, 515

\bibitem[Dekel, Bertschinger, \& Faber 1990]{}
Dekel, A., Bertschinger, E., \& Faber, S. M. 
1990, \apj, 364, 349

\bibitem[Dekel 1994]{}
Dekel, A. 1994, \araa, 32, 371

\bibitem[Ebeling et al. 1996a]{}
Ebeling, H., Allen, S. W., Crawford, C. S., Edge, A. C., 
Fabian, A. C., B\"ohringer, H., Voges, W., \& Huchra, J. P.,
1996a, in {\it R\"ontgenstrahlung from the Universe}, in press

\bibitem[Ebeling et al. 1996b]{}
Ebeling, H., Voges, W., B\"ohringer, H., Edge, A. C., 
Huchra, J. P., \& Briel, U. G.
1996b, \mnras, 281, 799

\bibitem[Eke, Cole, \& Frenk 1996]{}
Eke, V. R., Cole, S., \& Frenk, C. S. 
1996,\mnras, submitted

\bibitem[Fisher et al. 1994]{}
Fisher, K. B., Davis, M., Strauss, M. A., Yahil, A., 
\& Huchra, J. P. 1994, \mnras, 267, 927

\bibitem[Freedman 1996]{}
Freedman, W. L. 1996,
in {\it Critical Dialogues in Cosmology}, ed. N. Turok 
(Singapore: World Scientific), in press (astro-ph/9612024)

\bibitem[Henry \& Arnaud 1991]{}
Henry, J. P. \& Arnaud, K. A. 1991, \apj, 372, 410

\bibitem[Jones \& Forman 1984]{}
Jones, C. \& Forman, W. 1984, \apj, 276, 38

\bibitem[Kitayama \& Suto 1996]{}
Kitayama, T. \& Suto, Y. 1996, \apj, 469, 480

\bibitem[Kitayama \& Suto 1997]{}
Kitayama, T. \& Suto, Y. 1997, \apj, submitted

\bibitem[Kolatt \& Dekel 1997]{}
Kolatt, T. \& Dekel, A. 1997, \apj, in press

\bibitem[Lacey \& Cole 1993]{}
Lacey, C. \& Cole, S. 1993, \mnras, 262, 627

\bibitem[Liddle et al. 1995]{}
Liddle, A. R., Lyth, D. H., Schaefer, R. K., Shafi, Q., 
\& Viana, P. T. P. 1995, \astroph/9511057

\bibitem[Liddle et al. 1996]{}
Liddle, A. R., Lyth, D. H., Roberts, D., 
\& Viana, P. T. P. 1996, \mnras, 278, 644

\bibitem[Lilje 1992]{}
Lilje, P. B. 1992, \apj, 386, L33

\bibitem[Lubin et al. 1996]{}
Lubin, M. L., Cen, R., Bahcall, N. A., \& Ostriker, J. P.
1996, \apj, 460, 10

\bibitem[Ma 1996]{}
Ma, C.-P. 1996, \apj, 471, 13

\bibitem[Persic \& Salucci 1992]{}
Persic, M. \& Salucci, P. 1992, \mnras, 258, 14p

\bibitem[Press \& Schechter 1974]{}
Press, W. H. \& Schechter, P. 1974, \apj, 187, 425

\bibitem[Rugers \& Hogan 1996]{}
Rugers, M. \& Hogan, C. J. 1996, \apj, 459, L1

\bibitem[Tytler, Fan, \& Burles 1996]{}
Tytler, D., Fan, X. M., \& Burles, S. 1996, \nat, 381, 207

\bibitem[Viana \& Liddle 1996]{}
Viana, P. T. P. \& Liddle, A. R. 1996 \mnras, 281, 323

\bibitem[Walker et al. 1991]{}
Walker, T. P., Steigman, G., Schramm, D. N., Olive, K. A., 
and Kang, H.-S. 1991, \apj, 376, 51

\bibitem[Wampler 1996]{}
Wampler, E. J. 1996, \nat, in press

\bibitem[White, Efstathiou, \& Frenk 1993a]{}
White, S. D. M., Efstathiou, G., \& Frenk, C. S. 
1993a, \mnras, 262, 1023

\bibitem[White et al. 1993b]{}
White, S. D. M., Navarro, J. F., Evrard, A. E., 
\& Frenk, C. S. 1993b, \nat, 366, 429

\bibitem[Willick et al. 1996]{}
Willick, J. A., Strauss, M, A., Dekel, A., \& Kolatt, T.
1996, \apj, submitted (\astroph/9612240)

\bibitem[Willick et al. 1997]{}
Willick, J. A., Courteau, S., Faber, S. M., 
Burstein, D., Dekel, A., \& Strauss, M. A.
1997, \apjs, in press (\astroph/9610202)

\end{thebibliography}
\end{document}